\documentclass[prl,reprint,showpacs,floatfix]{revtex4-1}
\usepackage{graphicx,color,graphics,amsmath}

\begin{document}

\title{Creating spin-one fermions in the presence of artificial 
spin-orbit fields: \\
Emergent spinor physics and spectroscopic properties}

\author{
Doga Murat Kurkcuoglu and C. A. R. S{\'a} de Melo
}

\affiliation{
School of Physics, Georgia Institute of Technology, Atlanta, 
Georgia 30332, USA
}

\date{February 3, 2015}

\begin{abstract}
We propose the creation and investigation of a system of spin-one 
fermions in the presence of artificial spin-orbit coupling, 
via the interaction of three hyperfine states of fermionic atoms 
to Raman laser fields. We explore the emergence of spinor physics 
in the Hamiltonian described by the interaction between light and atoms, 
and analyze spectroscopic properties such as dispersion relation,
Fermi surfaces, spectral functions, spin-dependent momentum distributions
and density of states. Connections to spin-one bosons
and SU(3) systems is made, as well relations to the Lifshitz transition
and Pomeranchuk instability are presented.
\end{abstract}

\pacs{03.75.Ss, 67.85.Lm, 67.85.-d}

\maketitle

%
%

The field of ultra-cold atoms has been a very prolific area of 
research with the experimental realization of several fundamental
theoretical ideas such as Bose-Einstein 
condensation (BEC)~\cite{bose-1924, einstein-1924}, 
the Mott-Insulator transition in the Bose-Hubbard
model~\cite{fisher-1989} and the evolution from BCS to BEC 
superfluidity~\cite{leggett-1980, nozieres-1985, sademelo-1993}.
Strong connections to standard condensed matter physics have 
been developed, specially in the case of optical lattices,
and some very unique situations have also emerged due to the 
ability to control the trapping of atoms with different hyperfine 
states~\cite{bloch-2008}. One of these special cases is the creation 
of spin-$1/2$ bosons, where only two internal or
hyperfine states of bosonic atoms with integer spin were trapped 
and investigated experimentally~\cite{wieman-1997,silvera-1998}. 
The existence of bosonic or fermionic 
atoms with large integer or half-integer spins which have interactions 
that are independent of the hyperfine states could lead to the
realization of ${\rm SU(N)}$ invariant Hamiltonians, as evidenced
experimentally in the case of Strontium (Sr) atoms~\cite{ye-2014}. 
The realization of such exotic situations is promoting 
the field of ultra-cold atoms beyond the stage of simulating known 
Hamiltonians from diverse areas of Physics to the stage of creating 
novel Hamiltonians, which have no direct counterpart in any 
area of Physics.  An important example is the unusual case of spin-$1/2$ bosons
in the presence of artificial spin-orbit coupling, which was created 
experimentally~\cite{spielman-2011,pan-2012} and its effects 
on Bose-Einstein condensation were studied 
thoroughly~\cite{ho-2011, stringari-2012, baym-2013, pan-2014}.

In this manuscript, we propose also another exotic case 
corresponding to the creation of spin-one fermions in the presence
of artificial spin-orbit coupling, instead of the traditional spin-$1/2$ 
fermion case that has been recently studied 
theoretically~\cite{chuanwei-2011, zhai-2011, hu-2011, han-2012, seo-2012} 
and experimentally~\cite{zhang-2012, zwierlein-2012, spielman-2013, zhang-2014}. 
A potential candidate for such a situation is the Fermi isotope of 
Potassium $( ^{40} {\rm K} )$, which possess several hyperfine states 
that can be trapped. However, other high spin Fermi atoms are also potential 
candidates, such as Ytterbium (Yt) or Strontium (Sr).
We envision a situation that only three hyperfine states of 
the Fermi atom are trapped, and assume that Raman beams are used 
to produce artificial spin-orbit coupling in the fermionic system 
via light-atom interactions. The possibility of 
trapping three hyperfine states of fermions has a direct connection to 
color superconductivity, as we can also view the three different hyperfine
states as different colors (red, green and blue), and by controlling the
interactions between atoms in different hyperfine states we could create
several types of paired states, such as, red-green, red-blue, 
and green-blue~\cite{sademelo-2008}. We can also relate this system
to multi-band materials by thinking of the hyperfine states as labelling 
different energy bands, and if interactions can be tunned to produce 
superfluidity, we can create a multi-band superfluid in the presence of
spin-orbit coupling. Thus, the creation of spin-one fermions is not 
in violation of the spin-statistics theorem~\cite{pauli-1940}, as the 
spin degrees of freedom truly correspond to pseudo-spin states (color or 
band index). However, such a system possesses interesting spinor physics 
and spectroscopic properties to be discussed next.

We consider fermionic atoms with three hyperfine states coupled 
via Raman processes between states $1$ and $2$ as well as $2$ and $3$, 
such that there is a net momentum transfer ${\bf Q}_{12}$ to state $1$
and $-{\bf Q}_{23}$ to state $3$, resulting in the light-atom 
Hamiltonian matrix
\begin{equation}
\label{eqn:light-atom-hamiltonian-matrix}
{\bf H}_{\rm LA} ({\bf k})
=
\begin{pmatrix}
\varepsilon_{1}({\bf k}) & \Omega_{12} & 0  \\
\Omega_{12}^* & \varepsilon_{2}({\bf k})  & \Omega_{23} \\
0 & \Omega_{23}^* & \varepsilon_{3} ({\bf k}) \\
\end{pmatrix},
\end{equation}
written in the rotating frame, where the $\ell^{th}$ state 
carries momentum ${\bf k} - {\bf k}_{\ell}$. 
Each diagonal element 
$
\varepsilon_{\ell} ({\bf k}) 
= 
({\bf k} - {\bf k}_\ell)^2/(2m) + \eta_\ell
$ 
is the sum of the kinetic energy 
$({\bf k} - {\bf k}_\ell)^2/(2m)$ of the $\ell^{th}$ hyperfine state
after the net momentum transfer ${\bf k}_\ell$ and internal 
energy $\eta_\ell$. The momenta ${\bf k}_{\ell}$ are 
${\bf k}_1 = {\bf Q}_{12}$, ${\bf k}_2 = {\bf 0}$ and 
${\bf k}_3 = -{\bf Q}_{23}$.
The terms $\Omega_{\ell m}$ are the Rabi frequencies
coupling of adjacent hyperfine states, which can be taken to be real
such that $\Omega_{12} = \Omega_{12}^*$ and 
$\Omega_{23} = \Omega_{23}^*$. We can define an energy reference
via the sum $\sum_\ell \eta_\ell = \eta$, in this case we 
can set $\eta_1 = -\delta$,
$\eta_2 = \eta$ and $\eta_3 = + \delta$. 

When the Raman beams form an 
arbitrary angle, momentum transfers can be chosen to be
${\bf k}_1 = k_T {\hat {\bf x}},$
${\bf k}_2 = {\bf 0},$
and 
${\bf k}_3 = -k_T {\hat {\bf x}},$
with $0 \le k_T \le 2 k_R$, 
where $k_R = 2\pi/\lambda$ 
is the recoil momentum, 
and $\lambda$ is the photon wavelength. 
Assuming that all Rabi frequencies are the same
$(\Omega_{12} = \Omega_{23} = \Omega)$ the Hamiltonian 
of Eq.~(\ref{eqn:light-atom-hamiltonian-matrix}) 
reduces to 

\begin{equation}
\label{eqn:light-atom-spin-matrix}
\begin{pmatrix}
\varepsilon_{0}({\bf k}) - h_z ({\bf k}) + b_z  & -h_x/\sqrt {2} & 0 \\
-h_x/\sqrt {2} & \varepsilon_{0}({\bf k})  & -h_x/\sqrt {2} \\
0 & -h_x/\sqrt {2} & \varepsilon_{0} ({\bf k}) + h_z ({\bf k}) + b_z \\
\end{pmatrix},
\end{equation}
where
$
\varepsilon_0 ({\bf k}) 
= 
{\bf k}^2/(2m) 
+ 
\eta
$
is a reference kinetic energy which is the same for all
hyperfine states, 
$
h_z ({\bf k}) 
= 
2k_T k_x/(2m) + \delta
$
is a momentum dependent Zeeman field along the $z$-direction,
which is transverse to the momentum transfer direction,
$h_x ({\bf k}) = -\sqrt{2} \Omega$ is the {\it spin-flip} (Rabi)
field, and $b_z = k_T^2/(2m) - \eta$ is the quadratic Zeeman term.
A similar Hamiltonian was created recently in the NIST group
for spin-one bosonic atoms~\cite{spielman-2015}. 

The light-atom Hamiltonian matrix displayed 
in Eq.~(\ref{eqn:light-atom-spin-matrix})
can be expanded in terms of a subset of the ${\rm SU(3)}$ Gell-Mann matrices 
that includes the identity ${\bf 1}$ and the spin-one angular momentum 
matrices ${\bf J}_x$, ${\bf J}_z$ and ${\bf J}_z^2$. In compact notation,
the expansion reads 
\begin{equation}
{\bf H}_{\rm LA} ({\bf k})
= 
\varepsilon_0 ({\bf k}) {\bf 1}
- h_x ({\bf k}) {\bf J}_x
- h_z ({\bf k}) {\bf J}_z
+
b_z {\bf J}_z^2. 
\end{equation}
Written in this form the light-atom Hamiltonian matrix can 
be interpreted as describing spin-one fermions in the presence of 
momentum dependent {\it magnetic} field components 
$h_x ({\bf k})$, $h_z ({\bf k})$ and a quadratic 
Zeeman shift parametrized by the coefficient $b_z$. 
Notice that when $b_z = 0$ the system reduces to a spin-one fermion 
in the presence of a momentum dependent magnetic field. 
In this case the eigenvalues are 
$
E_{\alpha} ({\bf k}) 
= 
\varepsilon_0 ({\bf k}) 
- 
m_{\alpha} 
\vert 
h_{\rm eff} ({\bf k})
\vert,
$
with $m_{\alpha} = \{ +1, 0, -1 \}$, 
where the effective momentum dependent magnetic field amplitude is
$
\vert 
h_{\rm eff} ({\bf k})
\vert
= 
\sqrt
{
\vert 
h_x({\bf k})
\vert^2 
+
\vert 
h_z({\bf k})
\vert^2 
}.
$
%

%
%

Using Cardano's method~\cite{cardano-2007}, 
the eigenvalues of this spin-one fermion Hamiltonian
can be obtained analytically from the zeros of the characteristic polynomial 
$
P (\omega) 
= 
{\rm det}
\left[ 
\omega {\bf 1} - {\bf H}_{\rm LA} ({\bf k})
\right],
$ 
but the general expressions are quite cumbersone.
Thus, we also obtain the eigenvalues $E_{\alpha} ({\bf k})$ 
by direct diagonalization of ${\bf H}_{\rm LA} ({\bf k})$ 
to validate the analytical results and order them such that
$E_1 ({\bf k}) > E_2 ({\bf k}) > E_3 ({\bf k}).$ 

\begin{figure} [t]
\includegraphics[width = 1.00\linewidth]{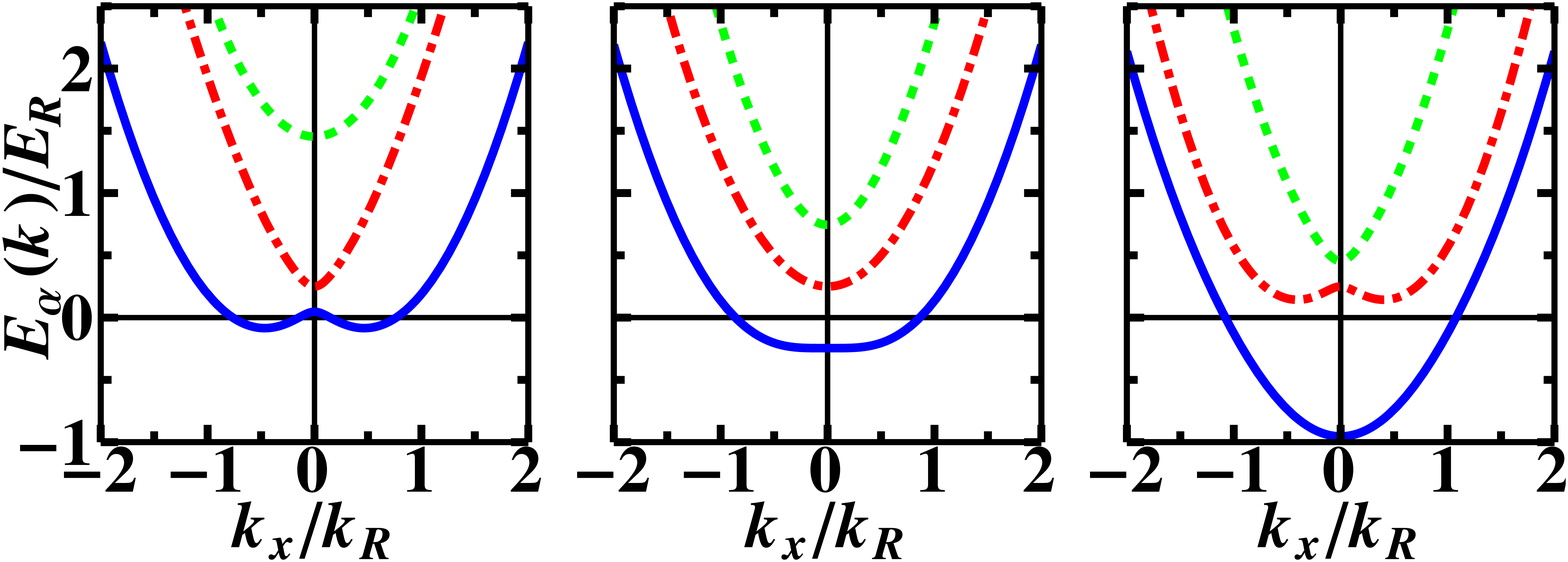}
\includegraphics[width = 1.00\linewidth]{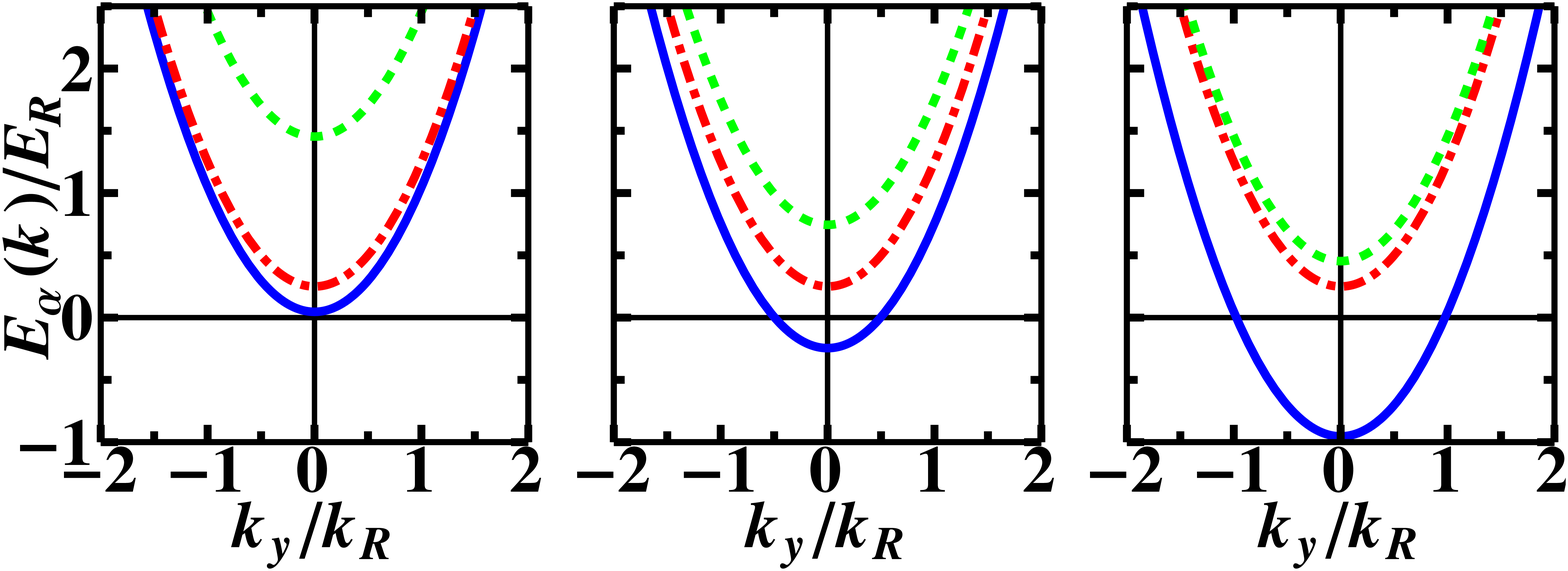}
\caption{
\label{fig:one}  
(color online) 
Eigenvalues $E_{\alpha} ({\bf k})$ in qualitatively different situations 
corresponding to momentum transfer $k_T = 0.5 k_R$, Rabi frequency 
$\Omega = 0.35E_R$ and quadratic Zeeman shift 
$b_z =  -E_R$ (left);
$b_z = 0$ (middle);
$b_z = E_R$ (right). The dashed-green line corresponds to  $E_1 ({\bf k})$,
the dot-dashed-red line to $E_2 ({\bf k})$, and the solid blue to
$E_3 ({\bf k})$. The top and bottom panels corresponds to cuts 
along the $(k_x, 0, 0)$ and $(0, k_y, 0)$ directions, respectively.
}
\end{figure}

In Fig.~{\ref{fig:one}}, we show plots of 
eigenvalues $E_{\alpha} ({\bf k})$ 
in qualitatively different situations corresponding
to momentum transfer $k_T = 0.5 k_R$, Rabi frequency 
$\Omega = 0.35E_R$ and three different
values of the quadratic Zeeman shift 
$
b_z 
= 
\{ 
-E_R, 0, E_R 
\}.
$ 
Along the $k_x$ direction, notice that 
a double minimum is present in $E_3 ({\bf k})$
when $b_z < 0$, and that a double minimum appears in $E_2 ({\bf k})$, 
when $b_z > 0$, while $E_3 ({\bf k})$ is very flat near $k_x = 0$ 
and $E_2 ({\bf k})$ has a single minimum when $b_z$ = 0. 
If our system consisted of spin-one bosonic atoms, a phase
transition would take place between a BEC at finite and zero momentum
as $b_z$ is increased from negative to positive values.

%

%
%

Since we are dealing with spin-one fermions, we investigate next 
the Fermi surfaces that emerge due to light-atom interactions 
as a function of the control parameters $k_T$, $\Omega$ and $b_z$,
and make connections to Lifshitz and Pomeranchuk instabilities 
found in condensed matter physics. 
We define an effective Fermi momentum $k_F$ via 
the total particle density 
$
n 
= 
3 k_F^3
/
(6\pi^2),
$
where the factor of $3$ indicates the presence of three internal states 
which lead to the three bands of the many-fermion system.
We also define the effective Fermi energy as $E_F = k_F^2/(2m)$
and make plots of Fermi surfaces are made assuming a   
density of $n = 10^{14} {\textrm{atoms}}/\textrm{cm}^3$. 

\begin{figure} [b]
\includegraphics[width = 1.00\linewidth]{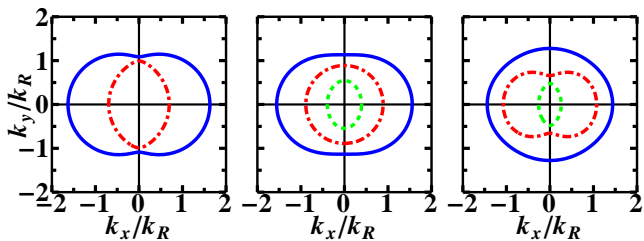}
\caption{
\label{fig:two}  
(color online) 
Fermi surfaces are shown in
qualitatively different situations corresponding 
to momentum transfer $k_T = 0.5 k_R$, Rabi frequency 
$\Omega = 0.35E_R$ and three different values of the quadratic Zeeman shift 
$b_z =  -E_R$ (left);
(b) $b_z = 0$ (middle);
(c) $b_z = E_R$ (right).
The values of the chemical potential are 
$\mu = 1.23 E_R$ (left), 
$\mu = 1.04 E_R$ (middle),
$\mu = 0.68 E_R$ (right) 
for particle density $n = 10^{14} {\textrm{atoms}}/\textrm{cm}^3.$ 
}
\end{figure}

In Fig.~\ref{fig:two}, 
we illustrate qualitatively different situations corresponding 
to $k_T = 0.5 k_R$, $\Omega = 0.35E_R$ and  
$b_z = \{-E_R, 0, E_R \}.$
Notice that in the middle panel of Fig.~\ref{fig:two} 
there is no quadratic Zeeman shift $(b_z = 0)$, 
but $k_T$ and $\Omega$ are non-zero. 
As described above, this implies that new fermionic bands 
$
E_{\alpha} ({\bf k}) 
= 
\varepsilon_0 ({\bf k}) 
- 
m_{\alpha}
\vert 
h_{\rm eff} ({\bf k})
\vert,
$
with $m_{\alpha} = \{ +1, 0, -1 \}$, 
emerge from three degenerate bands $\varepsilon_0 ({\bf k})$.
As a result, identical spherical Fermi surfaces 
associated with $\varepsilon_0 ({\bf k})$ become 
non-degenerate since the new energy 
dispersions are controlled by $\vert h_{\rm eff} ({\bf k})\vert,$ which 
is a function of $k_T$ and $\Omega$. With the exception of the
central band $E_2 ({\bf k})$, which still produces a spherical 
Fermi surface, the other two bands possess anisotropic Fermi surfaces
due to $\vert h_{\rm eff} ({\bf k})\vert$.

These effects are reminiscent of the Pomeranchuk~\cite{pomeranchuk-1958} 
instability in condensed matter physics, where deformations
in Fermi surfaces may emerge spontaneously in systems with anisotropic 
density-density interactions, without violating Luttinger's 
theorem~\cite{luttinger-1960}. In such cases, the 
resulting interactions produce deformations in the Fermi surfaces 
of the system, making them incompatible with the underlying symmetry
of the crystal. The easiest way to see this connection is to analyze 
the toy Hamiltonian 
$$
H 
= 
\sum_{{\bf k}, \alpha} 
\left[
\varepsilon ({\bf k}) {\hat n}_{\alpha} ({\bf k})
\right]
+
\frac{1}{2}\sum_{{\bf k}, {\bf k}^\prime \alpha \beta} 
F_{\alpha \beta}( {\bf k},{\bf k}^\prime) 
{\hat n}_{\alpha} ({\bf k}) {\hat n}_{\beta} ({\bf k}^\prime)
$$
where 
$
{\hat n}_{\alpha} ({\bf k})
=
c_{\alpha}^\dagger ({\bf k}) 
c_{\alpha}({\bf k})
$
is the number operator for spin $\alpha$.
The replacement of 
$
{\hat n}_{\alpha} ({\bf k}) 
=
\langle
{\hat n}_{\alpha} ({\bf k}) 
\rangle
+
\delta {\hat n}_{\alpha} ({\bf k} 
$
leads to the mean-field Hamiltonian
$
H 
= 
\sum_{{\bf k}, \alpha} 
\left[
E_{\alpha} ({\bf k})
{\hat n}_{\alpha} ({\bf k})
\right]
+
C.
$
The energy for internal state $\alpha$ is
$
E_{\alpha} ({\bf k})
= 
\varepsilon ({\bf k}) 
- 
h_{\alpha} ({\bf k}).
$
where 
$
\varepsilon ({\bf k}) 
=
{\bf k}^2/(2m),
$
is the kinetic energy of fermions of
mass $m$, and 
$ 
h_{\alpha} ({\bf k})
=
-
\sum_{\beta, {\bf k}^\prime}
\left[
F_{\alpha\beta} ({\bf k}, {\bf k}^\prime)
+
F_{\beta\alpha} ({\bf k}^\prime, {\bf k}) 
\right]
\langle
{\hat n}_{\beta} ({\bf k}^\prime)
\rangle,
$
is the {\it effective} field affecting the
$\alpha$-band. Lastly, the constant energy reference is
$
C 
=
\frac{1}{2}\sum_{{\bf k}, {\bf k}^\prime \alpha \beta} 
F_{\alpha \beta}( {\bf k},{\bf k}^\prime) 
\langle
{\hat n}_{\alpha} ({\bf k}) 
\rangle
\langle
{\hat n}_{\beta} ({\bf k}^\prime)
\rangle.
$
Notice that when $h_{\alpha} ({\bf k})$ does not have spherical symmetry,
then the Fermi surface for state $\alpha$ is deformed. 

In Fig.~\ref{fig:two}, a clear signature of the Pomeranchuk-like 
instability can be seen for the band with energy $E_2 ({\bf k})$ 
shown as the red dot-dashed line. However, notice 
that for fixed $k_T$ and $\Omega$, what drives the Fermi surface 
deformations is the quadratic Zeeman coupling $b_z$, 
that is, the $J_z$-$J_z$ spinor coupling instead of the density-density 
interactions. When $b_z = 0$, the Fermi surface corresponding
to $E_2 ({\bf k})$ is spherically symmetric, however when $b_z > 0$
$(b_z < 0)$ this Fermi surface suffers a predominant 
deformation along the $k_x$ $(k_y)$ direction. The Ising-Nematic 
order parameter 
$
{\cal N}_2 
= 
\int d{\bf k}
\left[
 k_y^2 + k_z^2 - 2 k_x^2
\right] 
\langle
\phi_2^\dagger ({\bf k})
\phi_2 ({\bf k})
\rangle
$
becomes zero for $b_z = 0$, positive for $b_z < 0$ and
negative for $b_z > 0$, where $\phi_2^\dagger ({\bf k})$ is
the creation operator for eingestate 2.
Similar Pomeranchuk-type deformations occur
for $E_1 ({\bf k})$ or $E_3 ({\bf k})$, however 
deformations are already present even for $b_z = 0$, 
because the spin-orbit coupling 
contains non-spherically-symmetric contributions 
through the effective field $h_{\rm eff} ({\bf k}).$

We also mention in passing the existence of a Lifshitz 
transition~\cite{lifshitz-1960}, 
which for fixed momentum transfer ${\bf k}_T$ and particle density
$n$, can be tuned via the Rabi frequency $\Omega$ and the quadratic
Zeeman coupling $b_z$. In Fig.~\ref{fig:two}, one can 
see a Lifsthitz transition for fixed $\Omega$ and changing $b_z$, 
as three Fermi surfaces
(genus 3) for $b_z = 0$ are reduced to two Fermi surfaces 
(genus 2) for $b_z = -E_R$. A phase diagram can be constructed mapping 
out these topological changes in the $\Omega$ versus $b_z$ plane. 

The effects of artificial spin-orbit 
and quadratic Zeeman coupling, due to light-atom interactions via the 
Raman scheme, can be further explored by 
investigating the three-component spinor wavefunctions.
For this purpose, we write the Hamiltonian as
\begin{equation}
{H}_{\rm LA}
= 
\sum_{\bf k}
{\bf \Psi}_{\bf k}^{\dagger} 
{\bf H}_{\rm LA} ({\bf k}) 
{\bf \Psi}_{\bf k},
\end{equation}
where ${\bf \Psi}_{\bf k}$ is a three-component spinor with
$
{\bf \Psi}_{\bf k}^{\dagger} 
= 
\left(
\psi_{1}^{\dagger} ({\bf k}),
\psi_{2}^{\dagger} ({\bf k}),
\psi_{3}^{\dagger} ({\bf k})
\right),
$
where $\psi_{s}^{\dagger} ({\bf k})$ 
represents the creation of a fermion in spin state $s$. 
When $s = 1$, the atom has momentum ${\bf k} - {\bf k}_T$
and $m_1 = +1$; when $s = 2$, the atom has momentum
${\bf k}$ and $m_2 = 0$; and when $s = 3$, the
atom has momentum ${\bf k} + {\bf k}_T$
and $m_3 = -1$.

The Hamiltonian $H_{\rm LA}$ can be diagonalized via the rotation 
$
{\bf \Phi} ({\bf k}) 
= 
{\bf U}({\bf k}) 
{\bf \Psi} ({\bf k}),
$
which connects the three-component spinor ${\bf \Psi} ({\bf k})$ 
in the original spin basis to the three-component spinor 
${\bf \Phi} ({\bf k})$ 
representing the basis of eigenstates. The matrix ${\bf U}({\bf k})$ 
is unitary and satisfies the relation 
${\bf U}^\dagger ({\bf k}) {\bf U} ({\bf k}) = {\bf 1}$. 
The diagonalized Hamiltonian is
$
{\bf H}_D({\bf k}) 
= 
{\bf U}({\bf k})
{\bf H}_{\rm LA} 
{\bf U}^{\dagger} ({\bf k})
$
with matrix elements 
$
\left[ 
{\bf H}_D({\bf k}) 
\right]_{\alpha \beta}
=
E_{\alpha} ({\bf k}) \delta_{\alpha \beta},
$ 
where $E_{\alpha} ({\bf k})$ are the eigenvalues of 
${\bf H}_{\rm LA} ({\bf k})$ discussed above.
The three-component spinor in the eigenbasis is  
$
{\bf \Phi}^{\dagger} ({\bf k}) 
= 
\left(
\phi^{\dagger}_{1} ({\bf k}),
\phi^{\dagger}_{2} ({\bf k}),
\phi^{\dagger}_{3} ({\bf k}),
\right),
$ 
where $\phi^{\dagger}_{\alpha} ({\bf k})$ is the creation 
operator of a fermion with eigenenergy $E_{\alpha} ({\bf k})$.
The unitary matrix
\begin{eqnarray}
{\bf U}({\bf k}) 
= 
\left(
\begin{array}{c c c}
u_{11} ({\bf k}) & u_{12} ({\bf k}) & u_{13} ({\bf k}) \\
u_{21} ({\bf k}) & u_{22} ({\bf k}) & u_{23} ({\bf k}) \\
u_{31} ({\bf k}) & u_{32} ({\bf k}) & u_{33} ({\bf k})  
\end{array}
\right)
\end{eqnarray}
has rows that satisfy the normalization condition
$
\sum_{s}
\vert u_{\alpha s} ({\bf k}) \vert^2 
= 
1.
$

Using a Stern-Gerlach technique, another spectroscopic property that 
can be measured is the spin-dependent momentum distribution 
\begin{equation}
\label{eqn:momentum-distribution}
n_{s} ({\bf k}) 
=
\sum_{\alpha} 
\vert u_{\alpha s} ({\bf k}) \vert^2 f [E_{\alpha}({\bf k})]. 
\end{equation}
We can fix the average number of particles 
$
N_s = \sum_{\bf k} n_s ({\bf k})
$
in each state $s$ independently, in which case chemical 
potentials $\mu_s$ for each state $s$ are necessary. 
However, when the total average number of particles 
$
N 
= 
\sum_{s} N_s
=
\sum_{s, \alpha}
\vert u_{\alpha s} ({\bf k}) \vert^2 
f [E_{\alpha}({\bf k})]
$ 
is fixed, we need only one chemical potential $\mu$.
The use of the normalization condition
$
\sum_{s} \vert u_{\alpha s} ({\bf k}) \vert^2 
= 
1
$ 
leads to 
$
N  
=  
\sum_{\alpha} f [E_{\alpha}({\bf k})].
$

\begin{figure}[t]
\includegraphics[width = 1.00\linewidth]{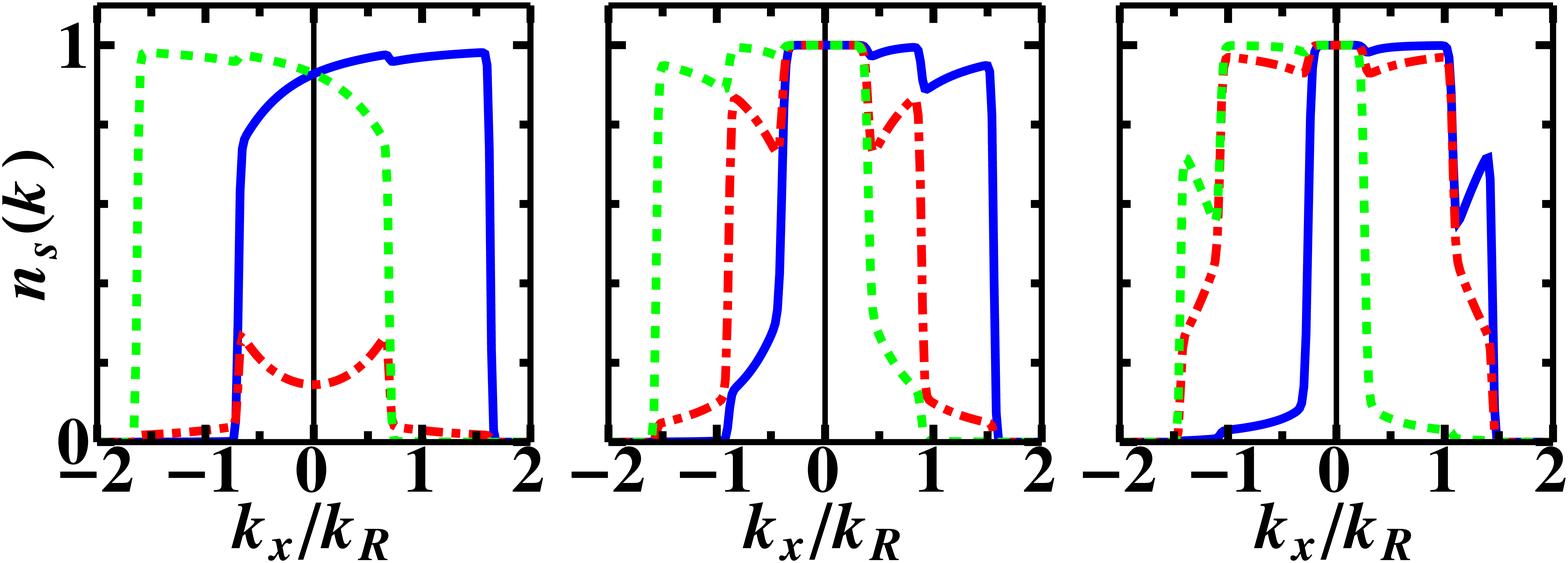}
\includegraphics[width = 1.00\linewidth]{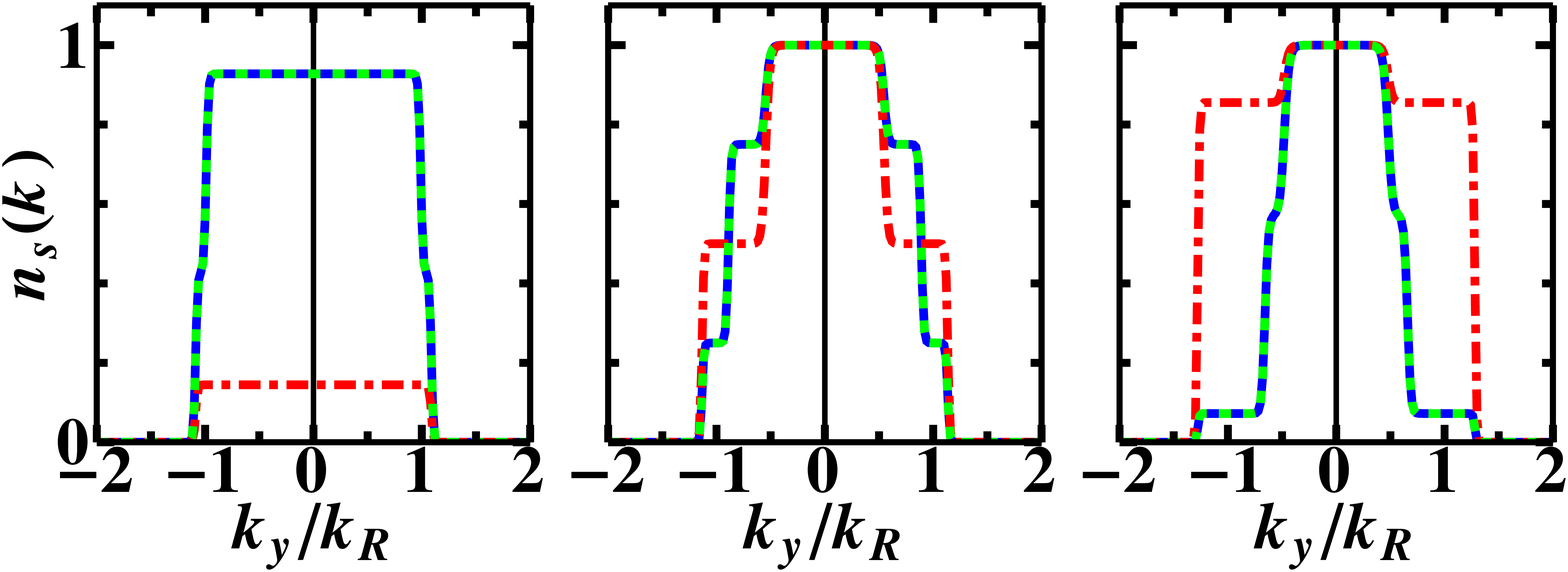}
\caption{
\label{fig:three}  
(color online) 
{Momentum distributions $n_s ({\bf k})$ for 
$s = 1$ $(m_1 = +1)$ (blue-solid curves), $s = 2$ $(m_2 = 0)$ 
(red-dot-dashed curves) and $s = 3$ $(m_3 = -1)$ (green-dashed curves),
with $T = 0.02 E_R \approx 0.01 E_F$.
The parameter values are
$b_z = -E_R$ (left panels), $b_z = 0$ (middle panels) and
$b_z = +E_R$ (right panels), with $k_T = 0.5 k_R$ and
$\Omega = 0.35 E_R$}.
}
\end{figure}

In Fig.~\ref{fig:three},
we show $n_s ({\bf k})$ at low temperatures 
for the simpler case where there is only one chemical potential.
The cross sections along $k_x$ with $k_y = k_z = 0$ are shown
in Fig.~\ref{fig:three} top panels, 
while the cross sections along $k_y$ with $k_x = k_z = 0$ 
are shown in Fig.~\ref{fig:three} lower panels.
In the top panels of Fig.~\ref{fig:three},
notice that $n_s ({\bf k})$ for states $s = 1$ $(m_1 = +1)$ 
and $s = 3$ $(m_3 = -1)$ do not have well defined parity, 
but are mirror images of each other. This is a reflection of 
the Hamiltonian invariance under the transformation 
$(k_x, m_1) \to (-k_x, m_3)$ and $(k_x, m_3) \to (-k_x, m_1)$. 

%

The momentum distributions shown in Fig.~\ref{fig:three} can be 
understood as follows.
The momentum transfer along the $k_x$ direction shifts the center 
of mass of the atom in state $s = 1$ with $m_1 = +1$ ($s = 3$ with
$m_3 = -1$) to be around $k_{T}$ $(-k_{T})$. While there is no 
momentum shift for the state $s = 2$ with $m_2 = 0$. 
In the limit of $\Omega \to 0$, $n_s ({\bf k})$ along $k_x$ have square shapes 
characteristic of degenerate fermions for each of the spin states.
However, momentum transfer can only occur when the lasers are on, 
which means $\Omega \ne 0$. This leads to mixing of the spin states 
and to a modification of the trivial momentum distributions 
via the coherence factors $\vert u_{\alpha s} ({\bf k}) \vert^2$. 
The dramatic effects of the coherence factors 
is seen on Fig.~\ref{fig:three} (top panels) where finite $\Omega$ 
causes strong deviations from square momentum 
distributions, due to the momentum-dependent mixing of different spin states.
However, $n_s ({\bf k})$ along the $k_y$ direction experience
no momentum transfer and are centered around zero.  
For $k_x = 0$, the light-atom Hamiltonian matrix is invariant 
under the transformations 
$(k_y, m_s) \to (-k_y, m_s)$, $(k_y, m_1) \to (-k_y, m_3)$, and 
$(k_y, m_3) \to (-k_y, m_1)$, such that the 
corresponding $n_s ({\bf k})$ along $k_y$ for states $s = 1$ and 
$s = 3$ are identical. The square like structures that emerge
are a consequence of the less dramatic dependence
of the coherence factors  $\vert u_{\alpha s} ({\bf k}) \vert^2$
on $k_y$. By symmetry, the same square structures also appear along 
the $k_z$ direction.

Notice that as $b_z$ increases
from negative to positive (left to right panels in Fig.~\ref{fig:three}),
$n_s ({\bf k})$ for state $s = 2$ along the $k_x$ and $k_y$ 
directions increase on average at fixed $\Omega$. 
This enhancement occurs because the energy of the $s = 2$ state becomes
increasingly lowered in comparison to the energy of the 
$s = 1, 3$ states, and spectral weight is transferred
from states $s = 1, 3$ to $s = 2$, causing a corresponding
decrease in the average $n_s ({\bf k})$ of the former states.
When $b_z$ becomes large and negative, 
the central state $(s = 2)$ is pushed up in energy with respect to the
$s = 1, 3$ states, and for densities such that the Fermi energy crosses 
only the two lowest states $(s = 1, 3)$, the system 
reduces to effective spin-$1/2$ fermions. However, 
when $b_z$ becomes large and positive, the central state $(s = 2)$ 
is pushed down in energy with respect to the $s = 1, 3$ states, and 
for densities such that the Fermi energy only crosses the $s = 2$ state, 
the system reduces to effective spin-zero (spinless) fermions.

\begin{figure}[htb]
\includegraphics[width = 1.00\linewidth]{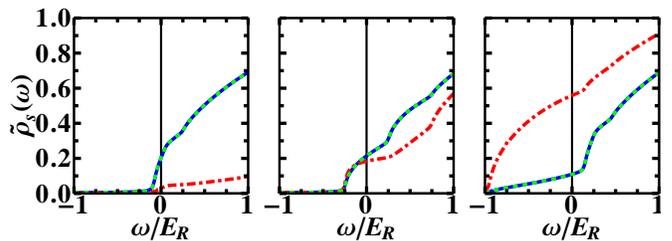}
\caption{
\label{fig:four}  
(color online) 
{Density of states ${\widetilde \rho}_s (\omega)
= \rho_s (\omega) E_F/N$ 
for $s = 1$ $(m_1 = +1)$ (blue-solid lines), 
$s = 2$ $(m_2 = 0)$ (red-dot-dashed lines),
and
$s = 3$ $(m_3 = -1)$ (green-dashed lines).
$N$ is the total number of particles and $E_F$ 
is the effective Fermi energy. We use a 
line-width broadening $\delta_{\ell w} = 0.01 E_R$. 
The parameters used are
$b_z = -E_R$ (left panel), $b_z = 0$ (middle panel) and
$b_z = +E_R$ (right panel), with $k_T = 0.5 k_R$ and
$\Omega = 0.35 E_R$ and $E_F = 1.95 E_R$.}
}
\end{figure}

The last spectroscopic quantity that we analyze is the spin-dependent
density of states (DOS)
\begin{equation}
\label{eqn:density-of-states}
\rho_{s} (\omega) 
= 
\sum_{{\bf k}, \alpha} 
\vert u_{\alpha s} ({\bf k}) \vert^2
\delta (\omega - E_\alpha ({\bf k})).
\end{equation}
Below the minimum of $E_3 ({\bf k})$ 
there are no states available,
that is, $\rho_{s} (\omega) = 0$ for 
$\omega \le \omega_{*} (\Omega, b_z, k_T) = {\rm min}_{\bf k} E_3 ({\bf k})$. 
The spin-dependent DOS for $\Omega = 0.35E_R$ and 
$b_z = \{ -E_R, 0, E_R \}$
are shown in Fig.~\ref{fig:four}. Notice that for $b_z = -E_R$ (left panel)
the spin-dependent DOS is non-zero only when 
$\omega \ge -0.09E_R$
and that for small values of $\gamma = (\omega - \omega_{*})/E_R$, 
the main contributions to the total DOS
$\rho (\omega) = \sum_s \rho_s (\omega)$ come from 
states $s = 1, 3$. In addition, for $b_z = 0$ 
(central panel), $\rho_s (\omega) \ge 0$ when 
$\omega \ge -0.27E_R$,
and the DOS for each spin component are 
comparable for small values of $\gamma$.
However, for $b_z = +E_R$ (right panel),
$\rho_s (\omega) \ge 0$ when $\omega \ge -1.00E_R$,
and the main contribution to $\rho (\omega)$ 
comes from $\rho_2 (\omega)$ for small values of 
$\gamma$, as state $s = 2$ has the lowest energy. 

In conclusion, we have proposed the creation of spin-one fermions
in the presence of spin-orbit fields and quadratic Zeeman shifts 
induced by light-atom interactions using a Raman coupling scheme. 
By adjusting the quadratic Zeeman shift, we have shown that we can tune 
the system from spin-zero to spin-one to spin-$1/2$ fermions. 
We have analyzed Lifshitz and Pomeranchuk instabilities 
for varying quadratic Zeeman shifts and studied several spectroscopic 
properties including energy dispersion, Fermi surfaces, spectral function, 
spin-dependent momentum distribution and density of states.

\newpage

\begin{figure}[htb]
\includegraphics[width = 2.00\linewidth]{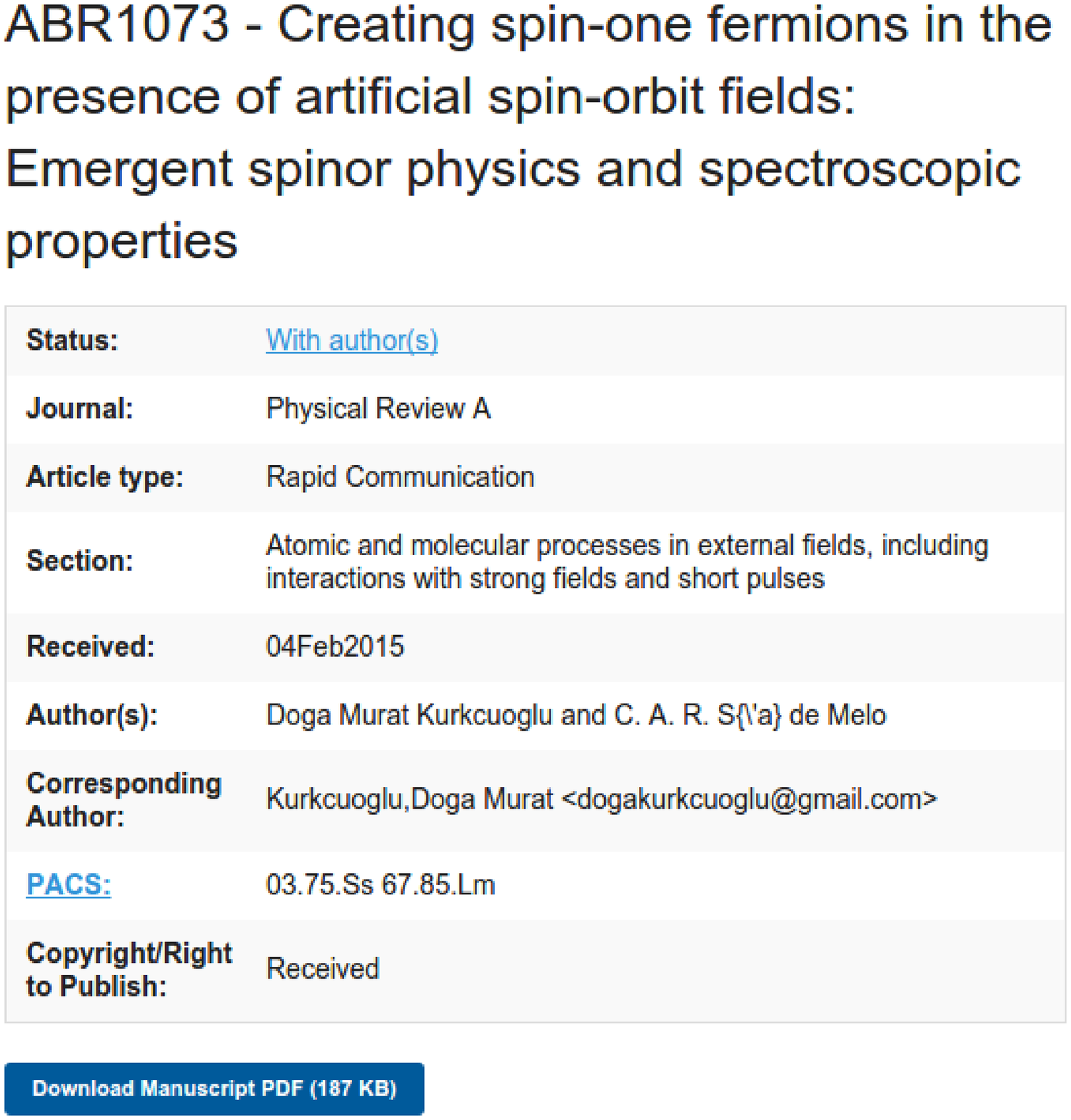}
\end{figure}

\end{document}